\documentclass[10pt]{svproc}

\usepackage{fixltx2e}

\usepackage{mathptmx}

\usepackage{microtype}

\usepackage{graphicx}

\usepackage[dvipsnames,svgnames,x11names]{xcolor}

\usepackage{url}
\usepackage[nocompress]{cite}

\usepackage{paralist}

\usepackage{sectsty}
\allsectionsfont{\normalsize\raggedright}
\subsectionfont{\normalsize\raggedright}

\begin{document}

\mainmatter

\title{Interoperable Convergence of Storage, Networking, and Computation\thanks{This manuscript was accepted to the FICC 2019 conference in San Francisco, CA, March 14-15, 2019. More details available at http://saiconference.com/FICC2019. FICC 2019 proceedings will be published in Springer series "Lecture Notes in Networks and Systems" and submitted for indexing to ISI Proceedings, MetaPress, Google Scholar, and Springerlink.}}
\titlerunning{Interoperable Convergence of Storage, Networking, and Computation}

\author{%
Micah Beck%
\inst{1,2}
\and%
Terry Moore%
\inst{1}%
\and%
Piotr Luszczek%
\inst{1}%
\and%
Anthony Danalis%
\inst{1}%
}
\institute{%
Department of Electrical Engineering and Computer Science \\
University of Tennessee, Knoxville, TN 37996 \\
\email{mbeck@utk.edu,tmoore@icl.utk.edu,luszczek@icl.utk.edu,adanalis@icl.utk.edu}\\
\and
Dr.\ Beck is an Associate Professor at University of Tennessee, Knoxville. he is currently on  detail to\\the National Science Foundation in the Office of Advanced Cyberinfrastructure. The work discussed herein\\was completed prior to his government service and  does not reflect the views,\\conclusions, or opinions  of the National Science Foundation or of the U.S. Government.
}

\maketitle

\begin{abstract}
In every form of digital store-and-forward communication, intermediate
forwarding nodes are computers, with attendant memory and processing resources.
This has inevitably stimulated efforts to create a wide-area infrastructure
that goes beyond simple store-and-forward to create a platform that makes more general
and varied use of the potential of this collection of increasingly powerful
nodes. Historically, these efforts predate the advent of globally
routed packet networking. The desire for a converged infrastructure of this
kind has only intensified over the last 30 years, as memory, storage, and
processing resources have increased in both density and speed while
simultaneously decreasing in cost. Although there is a general consensus that
it should be possible to define and deploy such a dramatically
more capable wide-area platform, a great deal of investment in research
prototypes has yet to produce a credible candidate architecture.  Drawing on
technical analysis, historical examples, and case studies, we present an
argument for the hypothesis that in order to realize a distributed system with
the kind of convergent generality and deployment scalability that might qualify
as "future-defining," we must build it from a small set of simple,
generic, and limited abstractions of the low level resources (processing, storage and
network) of its intermediate nodes.
\end{abstract}

\begin{figure*}
\centering
\includegraphics[width=4in, viewport=50 150 750 500]{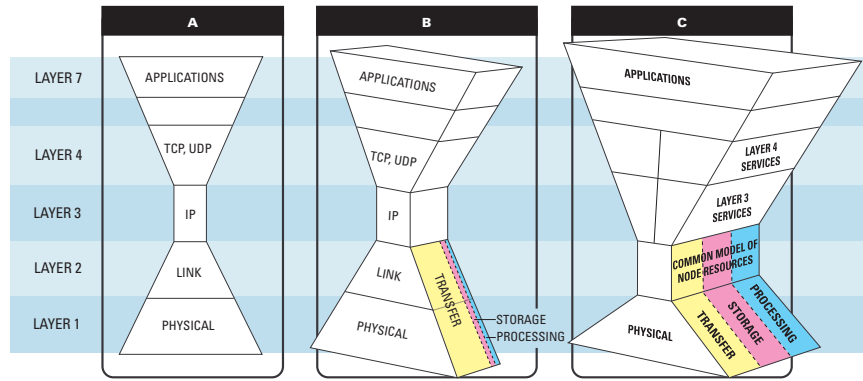}
\caption{The Hourglass vs. The Anvil}
\label{fig:anvil}
\end{figure*}

\twocolumn

\section{Introduction}
\label{sec:intro}

A variety of technological, economic, and social developments --- most notably
the general movement toward Smart Cities, the Internet of Things, and other
forms of ``intelligent infrastructure'' \cite{mynatt2017national} --- are
prompting calls from various quarters for something that the distributed
systems community has long aspired to create: A next-generation network
computing platform. For
example, the authors of a recent Computing Community Consortium white paper,
writing with the US ``Smart Cities'' initiative~\cite{nitrd_scc} in view,
express the research challenge as follows:

``What is lacking---and what is necessary to define in the future---is a
  common, open, underlying `platform', analogous to (but much more complex
  than) the Internet or Web, allowing applications and services to be developed
  as modular, extensible, interoperable components. To achieve the level of
  \textbf{interoperation} and innovation in Smart Cities that we have seen in
  the Internet will require \emph{federal investment in the basic research and
  development} of an analogous open platform for intelligent infrastructure,
  tested and evaluated openly through the same inclusive, open,
  consensus-driven approach that created [the]
  Internet.''~\cite{nahrstedt2017city} [Emphasis in source]

The experiences of the last two decades have made the distributed systems
community acutely aware of how elusive the invention of such a future-defining
platform is likely to be~\cite{Anderson05overcomingbarriers}. Achieving
this vision has been the
explicit or implicit ambition of a succession of well funded and energetically
pursued research and development efforts within or around this community,
including Active Networking~\cite{Tennenhouse96towardsan}, Grid
Computing~\cite{foster1999grid}, PlanetLab~\cite{Chun:2003:POT:956993.956995},
and GENI~\cite{GENIMcGeer16}, to name a few. Although these broad efforts have
produced both valuable research and useful software results, nothing delivered
so far has achieved the \textit{deployment scalability} necessary to
initiate the kind of viral growth that everyone expects such an aspirational
platform to exhibit. At the same time, chronic problems with network hotspots
were an early and persistent sign that the Internet's stateless, unicast
datagram service had scalability limitations with respect to data volume and/or
popularity. This fact has led to 
increasingly sophisticated and increasingly
expensive technology ``workarounds,'' from the FTP mirror sites and
Web cache hierarchies of the early years, to the content delivery networks
(CDN) and commercial Clouds we see today.

The central idea of this paper is that the appropriate common service on which
to base an interoperable platform to support distributed systems is an
abstraction of the low layer resources and services of the intermediate node,
i.e., a generalization of the Internet stack's layer 2. The ``Internet Convergence'' of
the 1990's developed the ``hourglass'' paradigm, with best-effort datagram
delivery as the common service, or ``spanning-layer,'' at its narrow
waist~\cite{clark1995interoperation};
we believe that the paradigm required by the data
saturated world now emerging in edge/fog environments is more accurately
pictured as an ``anvil'' (Figure~\ref{fig:anvil}),
with a common service interface that
exposes storage/buffer, network, and processor resources in a programmable way.
Drawing on technical
analysis and historical examples, we argue that in
order to build distributed systems with the kind of interoperability,
generality and deployment scalability that might qualify as
``future-defining,'' we must implement them using a small set of simple,
generic, and limited abstractions of the data transfer, storage and processing
services available at this layer. In our model, these abstractions all revolve
around the fundamental common resource, the memory/storage buffer.

\section{Background}
\label{sec:background}

Given the inclination of computer scientists to add features, the fact that
every form of digital store-and-forward communication (including the Internet)
has intermediate forwarding nodes that are computers, with attendant memory and
processing resources, makes attempts to create a wide area infrastructure with
services beyond  simple store-and-forward inevitable. Such efforts to make more
general use of these increasingly powerful nodes--- a \emph{generalized
converged network}, in our terminology---predate the advent of globally routed
packet networking (e.g. \texttt{uux}~\cite{uuxPOSIX}).  The exponentially
increasing density and speed, and rapidly decreasing cost of memory, storage
and processing resources over the past 30 years has only intensified the desire
to define and scalably deploy a converged infrastructure of this general
description. Yet despite the general consensus that it should be possible
to do so, this aspiration has remained unfulfilled.

One problem is that the goal of converged networking runs in the opposite
direction of the traditional architectural approach of the  Internet
design community, which insists that services other than datagram delivery must
be kept out of the Network Layer of the communication protocol stack.
This community maintains that the ability of the Internet to function properly and
to continue growing globally depends on keeping this common service layer
``thin'', in the sense that it provides services that are simple, generic and
limited.
From this point of view,
services other than datagram delivery should be implemented in systems
connected to the Internet as communication endpoints. Various rationales
supporting this point of view are collectively referred to as ``End-to-End
Arguments''\cite{Saltzer:1984:EAS:357401.357402}.

Since a router that has substantial system storage (i.e. other than network
buffers) and generalized computational resources (i.e. other than forwarding)
is neither difficult nor expensive to build, there have been numerous efforts
to resist this orthodox point of view. The simple fact that storage and
computational resources can be provisioned and located throughout the network
at reasonable cost stimulates efforts in this direction. Moreover, the apparent
opportunity to create such a powerful distributed infrastructure presents a
temptation that is inherently difficult for computer scientists and engineers
to resist. These facts, however, do not make it a good idea to add extensions
to the fundamental service of the global Internet, nor do they ensure that if
it is built, service creators and users will adopt it at a scale sufficient to
enable economic sustainability beyond the prototype stage.  Indeed, while a
number of plausible network service architectures have been defined that can
provide access to such distributed
resources~\cite{Tennenhouse96towardsan,rfc3234}, the widespread deployment
of extended services on a converged wide area infrastructure has proved elusive.

Perhaps an even more compelling reason for the continued drive to create such a
converged infrastructure is that some important distributed applications cannot
be efficiently and effectively implemented through decomposition into two
components, one implemented by a ``thin'' datagram delivery service
in the core of the network, and the
other implemented at ``fat'' endpoints.  For example, some applications require
an implementation that is sensitive to the location of storage and computation
in the network topology.  Point-to-multipoint communication was an early and
obvious example.  Using simple repetition of unicast datagram delivery was
viewed as too inefficient by early Internet architects, but an efficient tree
could be built only through the use of network topology information.  Such low
level information was seen as inappropriate for users of the ``thin'' and
stable Network layer to access.  Thus, multicast was added to Layer 3,
fattening that thin layer with services that seemed to address this issue.
However, IP multicast has proved difficult to standardize and has failed to
achieve the universal deployment of ``simple, generic and limited'' unicast IP
datagram delivery.

But problems with lack of generality in the intermediate nodes were manifest
even in highly successful Internet applications.  The early growth of the
Internet was fueled by applications that seemed to fit the unicast datagram
delivery model well enough: FTP and Telnet.  Of these, the one-to-many nature
of FTP, albeit asynchronous, created a problem in the distribution of popular
and high-volume files.  Ignoring the implications of  topology led to
ineffective use of the network, with hotspots at servers that attracted high
volumes of traffic and unnecessary loads placed on expensive and overburdened
wide area links.  The result was the creation and management of collections of
FTP mirror sites~\cite{DongarraNetlib08}, and the ubiquitous invitation for
users to ``choose a mirror site near you'', which meant 
the use of approximate information about network topology by the end-user, at a
level above even the Internet stack's Application Layer.

The advent of the World Wide Web exacerbated the problem of indiscriminate
access to servers with no reference to network topology or even geography.
Mirror sites for file download proliferated, and redundancy in the storage of
all high-traffic Web content became a necessity.  A Network layer that hides
topology from its clients is, after all, an inherently inadequate platform on which to build
high traffic globally distributed systems.  The need to work around this
reality gave rise to automated Web
caching~\cite{Chankhunthod95ahierarchical,Wessels98icpand} and server
replication~\cite{KirkpatrickIBMOlympics96,Beck98theinternet2}, which were
precursors to modern Content Delivery
Networks~\cite{BuyyaCDN08,Nygren10theakamai}.

It should be noted that although both Web caching and server replication are
obvious examples of the convergence of networking and storage, they also
require computation in the implementation of policy and server-side
processing; and so in fact they represent convergence of all three fundamental
computational resources. We examine the approach to convergence that they
represent in more detail in section~\ref{workaround} below.
Following a different strategy, Logistical Networking, discussed in
section~\ref{subsec:logistical}, implements a convergence 
of networking and storage service that avoids the need for general computation
by minimizing policy and other server-side
processing~\cite{Beck02anend-to-end}, but was later extended to include limited
server-side operations~\cite{Beck03anend-to-end}.

\section{The Convergence Spectrum}
\label{sec:convergence}

The interplay between technological divergence and  convergence is a dialectic
with a long history.  In the area of computing and communications, there was an
early divergence in the conception and implementation of several different
information technology resources.  Because of the phenomenon of path
dependence~\cite{david2007path}, such divergence has tended to be
self-reinforcing, leading to a set of familiar technology \emph{silos},
such as data transmission and broadcast using radio frequency signals,
virtual circuits, switches and gates and magnetic or solid state storage cells.
The success of the Internet in the 1990`s provided the
foundation for the substantial or partial convergence of various traditional
telecommunication silos--telephony, broadcast television, etc.--in this century~\cite{messerschmitt1996convergence},
but the three fundamental silos at the base of computing--storage, processing,
networking--have remained as entrenched as ever.
%




The early divergence of basic computational resources
has given rise to conceptual, technological and
organizational {\it silos} corresponding to isolated communities.
Formal models and methods of reasoning have been adapted to deal with the
complexity and specific issues of each niche. For example Boolean logic is a
useful model of solid state circuits, and "stateless" communication is a useful
model of a wide area data network built out of switches and FIFO line buffers.
%

\begin{figure}
\vspace*{-2ex}
\centering
\includegraphics[width=3in, viewport=0 250 500 490]{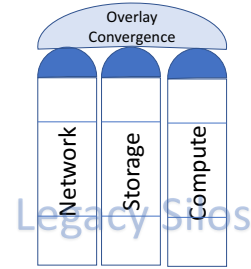}
\vspace*{-2ex}
\caption{Overlay Convergence of Legacy Silos}
\vspace*{-5ex}
\label{fig:silos}
\end{figure}

The development of silos has been an enabling strategy for modeling and
optimization of these quickly evolving technological fields.  However, they
have also led to the creation of service stacks with highly
specialized services at the top layers (see Figure~\ref{fig:silos}).  But
because the low level resources that these silos encapsulate can only be
accessed through high level services, this inevitably tends to create barriers
to the flexible and efficient use of constituent low level resources {\it in combination}.

The problem with silos as a strategy for dealing with the complexity and
specialization of disparate underlying technologies has become more pronounced
due to the evolution of low level systems toward general mechanisms that
utilize  processors or digital logic controlled by software, firmware or by
hardware designed using computerized tools.  Such generality in low level
mechanisms holds out the possibility of the implementation of highly efficient
system architectures, with optimizations that span traditionally disparate
resources.  The challenge is to bridge or eliminate the existing silos, or, in
other words, to implement \emph{convergence}.

We say that a service interface (i.e., an API) is {\em converged} if it gives
unified access to multiple low-level resources (or services) traditionally
available only through isolated service silos.
Historical examples of system design that leverage convergence include
the auto-increment register,  direct memory access I/O and vector processing.

When the goal is to achieve convergence for a service interface
using previously non-interoperable resources, there are two
fundamentally different ways to go about it: \emph{overlay
convergence} which combines silos at a layer above their high level services,
and \emph{interoperable convergence}, which strives to unify their foundations.
These two strategies lie at the ends of a spectrum along which a variety
of familiar examples can be arrayed.

Overlay convergence is the most common approach
lying at one end of the spectrum and creating
a high level interface that provides access to a number of traditionally
separate service silos.  We term this approach {\em overlay convergence}
because it typically involves the creation of a service that provides unified access to
the existing service silos from above, through their high level client interfaces (see
Figure~\ref{fig:silos}).  By contrast, at the other end of the convergence
spectrum is what we call {\em
interoperable convergence}. We say that a platform is interoperably converged if it minimizes the
imposition of unnecessary high-level structure or performance
costs when applying different low-level services, so that those underlying common resources
can be accessed without incurring the overhead and restrictions that
are associated with complex and specialized service silos.

Both overlay and interoperable forms of convergence seek to create a common
service, or {\it spanning layer}, which supports a generalized set of
applications requiring resources that were previously segregated. As
exemplified in the classic case---the ``narrow waist'' of the Internet's
``hourglass'' protocol architecture (see Figure~\ref{fig:anvil})---the purpose
of such a spanning layer is to enable interoperability in the support of this
rich category of applications~\cite{clark1995interoperation}.

Some examples that fall along this spectrum and illustrate these different approaches
include the following:

\begin{itemize}
\item The BSD kernel created an overlay convergence of Unix process and local
  file management with local and wide area networking through the addition of
  the {\em socket} related system calls.
While some calls that act on file descriptors such as {\tt read()} and {\tt write()}  were extended
to operate on sockets, the level of integration is mainly syntactic and does not
extend deeply into integration of common functions such as buffer management.

\item Following the implications of this example, in order to move data stored in a file to a TCP stream in UNIX, it was
  originally necessary to move it into a process' address space using the
  {\tt read()} system call and then inject it into the TCP stream using {\tt send()} (see Figure~\ref{fig:sendfile}). A
  more interoperable approach is a combined {\tt sendfile()} system call was
  added as an extension to Linux that allows data to be transferred from storage into a kernel
  memory buffer and from there directly to the network without moving it to
  process memory or using a dedicated network data buffer.
 However this buffer management solution is applicable only in quite specific scenarios.
 We thus characterize it as a {\em workaround}.
\item A distributed file system converges storage and data movement in a more
interoperable manner.
These resources are otherwise available only through local file management
    and networked file transfer tools.



\begin{figure}
\vspace*{-8ex}
\centering
\includegraphics[width=3.5in]{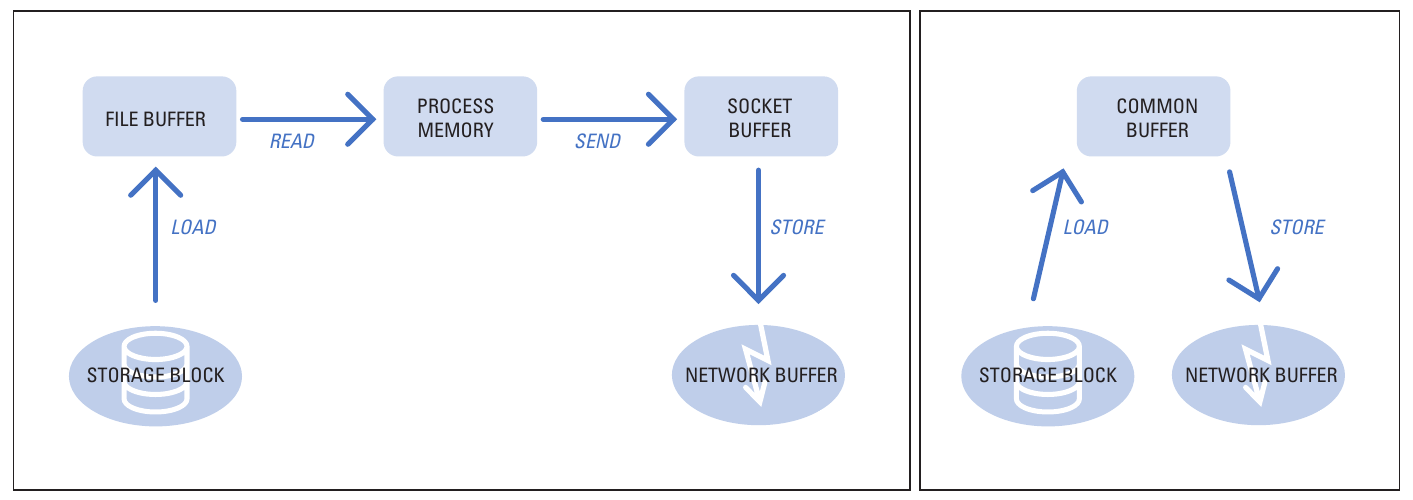}
\vspace*{-6ex}
\caption{Read-Send vs. Sendfile}
\vspace*{-5ex}
\label{fig:sendfile}
\end{figure}

\item A database system can store a set of tuples without order, but
traditional data movement tools operate on files. Thus, it is necessary to
serialize a set of tuples as a file in order to send it to a remote
database system. The file is transferred serially, using TCP with
retransmission to keep the serialized data in order. A somewhat
interoperable approach would generate the serialized stream representing
the tuple set on demand, rather than creating and storing it as a complete
file. A more interoperable approach would be to implement a specialized
protocol that takes advantage of the lack of natural sequentiality in the
tuple set to perform retransmission out-of-order. This might require
additional work to ensure that the new protocol was ``TCP-friendly'' when
used in public shared networks.

\item A data analysis system (such as
MapReduce~\cite{Dean:2008:MSD:1327452.1327492}) traditionally consists of a
deep data store and a
dedicated compute resource such as a cluster or a shared-memory parallel
computer. Visualization typically requires data to be moved from the data store to
the compute resource which then returns its results to the data store. User
access then requires that the visualization output be moved to and
interpreted by a human interaction system. A more interoperable approach
would allow computations to be applied to the data in the data store
(in-situ), and for the user to interact with the results of that
computation directly as it occurs.
\end{itemize}

\section{Deployment Scalability}
\label{sec:scalability}

\begin{figure}
\vspace*{-3ex}
\includegraphics[width=\linewidth]{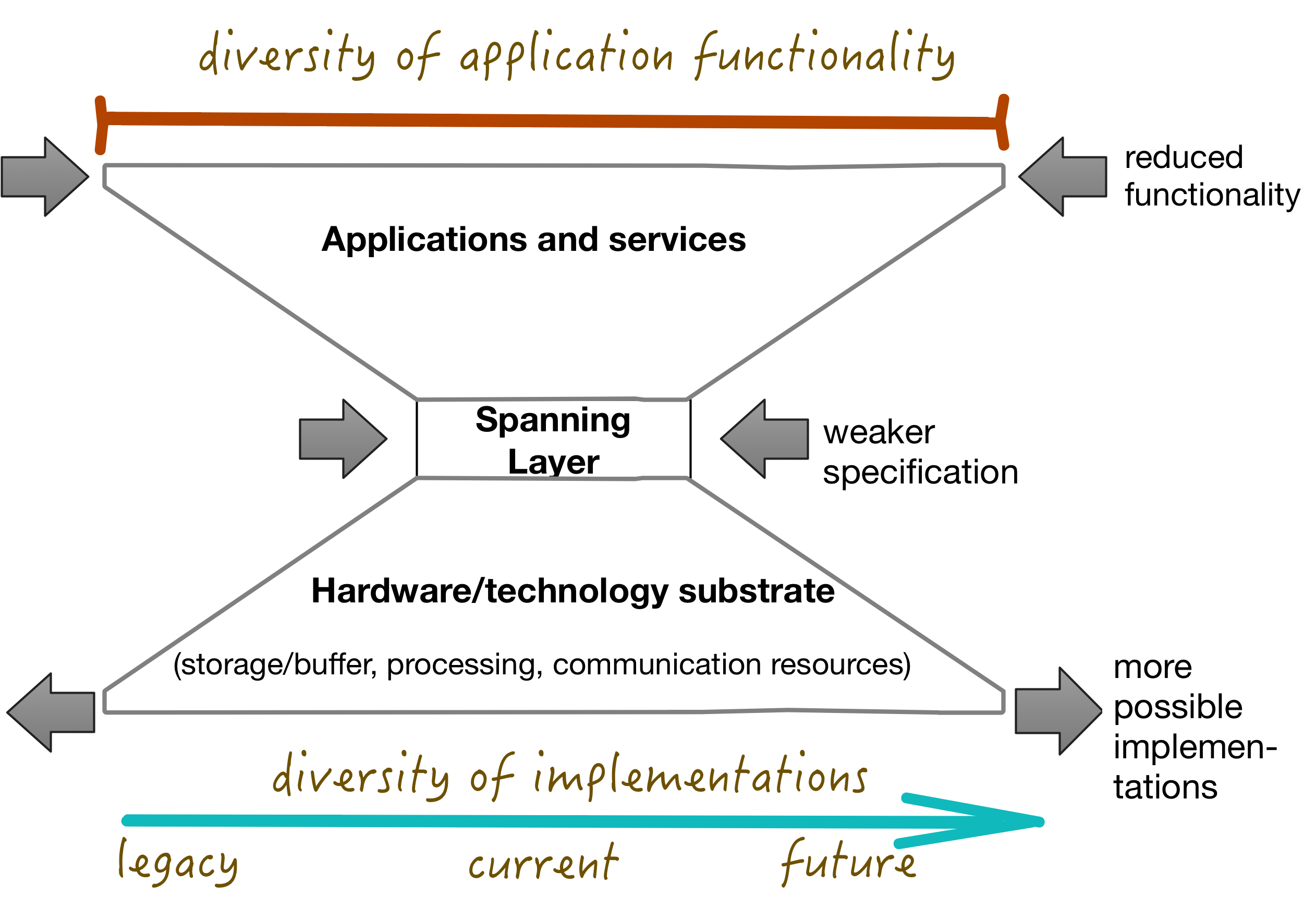}
\vspace*{-7ex}
\caption{The ``hourglass model'' for a system software stack. The goal is to achieve deployment scalability while maximizing the diversity of applications. The arrows indicate the design tensions involved in creating a spanning layer with a minimally sufficient service specification.
\vspace*{-5ex}
\label{f_spanning}}
\end{figure}

When it comes to creating layered software stacks for large and diverse
communities, the importance of a well designed spanning layer is difficult to
overestimate. Most critically, a successful hourglass design connects directly
with the diverse demands of \emph{multi-stakeholder ecosystems}. As shown in
Figure~\ref{f_spanning}, a well designed spanning layer can be implemented on
many different substrates, yielding a wide ``lower bell,'' and yet also support
an equally great variety of applications, yielding a wide ``upper bell.'' A
wide lower bell means that the spanning layer can be implemented on
heterogeneous hardware/software platforms, enabling applications and services
above the spanning layer to access and utilize these diverse resources. The
wider the lower bell, the stronger the assurance that both legacy and future
platforms will support many different stakeholders. A wide upper bell means
that the small set of primitive services that the spanning layer makes
available on the system's nodes can be composed, often using additional
stakeholder-provisioned resources, to support a broad diversity of higher-level
services and applications. The wider the upper bell, the stronger the assurance
that more specialized application communities can build what they need atop the
shared infrastructure. Thus, a successful hourglass architecture, with
capacious upper and lower bells, will lower or eliminate barriers to adoption
for a wide variety of stakeholders.

The spanning layer in such a successful hourglass-shaped stack is said to be
``narrow'' because ``\ldots it represents a minimal and carefully chosen set of
global capabilities that allows both higher-level applications and lower-level
communication technologies to coexist, share capabilities and evolve
rapidly~\cite{Peterson:2011:CNF:2012058}.'' But deliberately creating a
spanning layer with such a \textit{minimally sufficient} specification has
proven to be exceedingly difficult, as is shown by the list of major efforts
over the last two decades that have attempted to do so (see
Section~\ref{sec:intro}).

Rationally evaluating alternative strategies requires a criterion for success.
Accordingly, we introduce the concept of \emph{deployment scalability} as the
design goal of the foundational spanning layer of a converged infrastructure.
We define {\em deployment scalability} as \textit{widespread acceptance,
implementation and use of a service specification}. Since spanning layers
define communities of interoperability, the greater the deployment scalability
of a given spanning layer, the larger the community of interoperability it can
achieve.  The workarounds we have described build overlay converged systems,
but they cannot achieve a sufficient level of deployment scalability, which
constrains the size of the communities they can sustainably support.

In a recent paper~\cite{beck_hourglass_2018}, Beck makes an argument for a fundamental design
principle of the common service underlying systems that exhibit deployment scalability:
\begin{quote} {\bf The Deployment Scalability Tradeoff}
There is an inherent tradeoff
between the deployment scalability of a specification and the degree to which
that specification is weak, simple, general and resource limited.
\end{quote}

\noindent The terms ``simple, generic and resource limited'' are derived from the classic
paper ``End-to-End Arguments in System Design'' by Saltzer, Reed and Clark which
discusses them in the context of  Internet architecture.
The term ``weak'' refers to logical weakness of the service specification as a
theory of program logic, and is due to Beck's  partial formalization of the
arguments in that paper.
Stating this principle as a tradeoff  is a further refinement of the
usual interpretation of the original paper as an
absolute rule (or principle) requiring or prohibiting particular design
choices~\cite{RealE2e}.

The classic example of the application of the End-to-End Principle,
from which its name is derived, is the location of the detection of
data corruption or packet loss or reordering in the TCP/IP
stack~\cite{Saltzer:1984:EAS:357401.357402}. The scalability argument for end-to-end detection of faults is that
removing such functions from the spanning layer makes it weaker,
and therefore potentially admits more possible implementations.
Because fault detection can be implemented above the spanning layer,
the set of applications supported is not reduced.

The evolution of process creation in Unix teaches a similar lesson.
In early operating systems it was common for the creation of a new process to
be a privileged operation that could be invoked only from code running with
supervisory privileges.
There were multiple reasons for such caution,
but one was that the power to allocate operating system resources that
comprise a new process was seen as too great to be delegated to the
application level.
Another reason was that the power of process creation
(for example changing the identity under which the newly created process
would run) was seen as too dangerous.
This led to a situation in which command line interpretation was a
near-immutable function of the operating system that could only be changed by
the installation of new supervisory code modules,
often a privilege open only to the vendor or system administrator.

In Unix, process creation was reduced to the \texttt{fork()} operation,
a logically much weaker operation that did not allow any of the attributes of
the child process to be determined by the parent, but instead required that
the child inherit such attributes from the parent~\cite{Ritchie74theunix}.
Operations that changed sensitive properties of a process were factored out
into orthogonal calls such as \texttt{chown()} and \texttt{nice()},
which were fully or partially restricted to operating in supervisory mode;
and \texttt{exec()} which was not so restricted but which was later extended
with properties such as the \emph{setuid} bit that were implemented as
authenticated or protected features of the file system.
The decision was made to allow the allocation of kernel resources by
applications, leaving open the possibility of dynamic management of such
allocation by the kernel at runtime, and creating the possibility of
``Denial of Service'' type attacks that persists to this day.

These two classical examples of interoperable convergence point to a
significant issue.  Changing the low level services on which existing silos are
built requires the redesign and reimplementation of complex higher level
service stacks.  The influence of path dependent thinking and the pain of
abandoning "sunk investments" explain the natural tendency of service provider
communities to develop {\it workarounds} that preserve widely deployed lower
level services. In Section~\ref{workaround} below, we analyze some familiar
overlay workarounds to the problems that can be traced to the tradeoffs that
the designers of the Internet made.

\section{Web Caching and CDNs: A Case Study in Overlay Workarounds}
\label{workaround}

\begin{figure}
\vspace*{-7ex}
\centering
\includegraphics[width=3in]{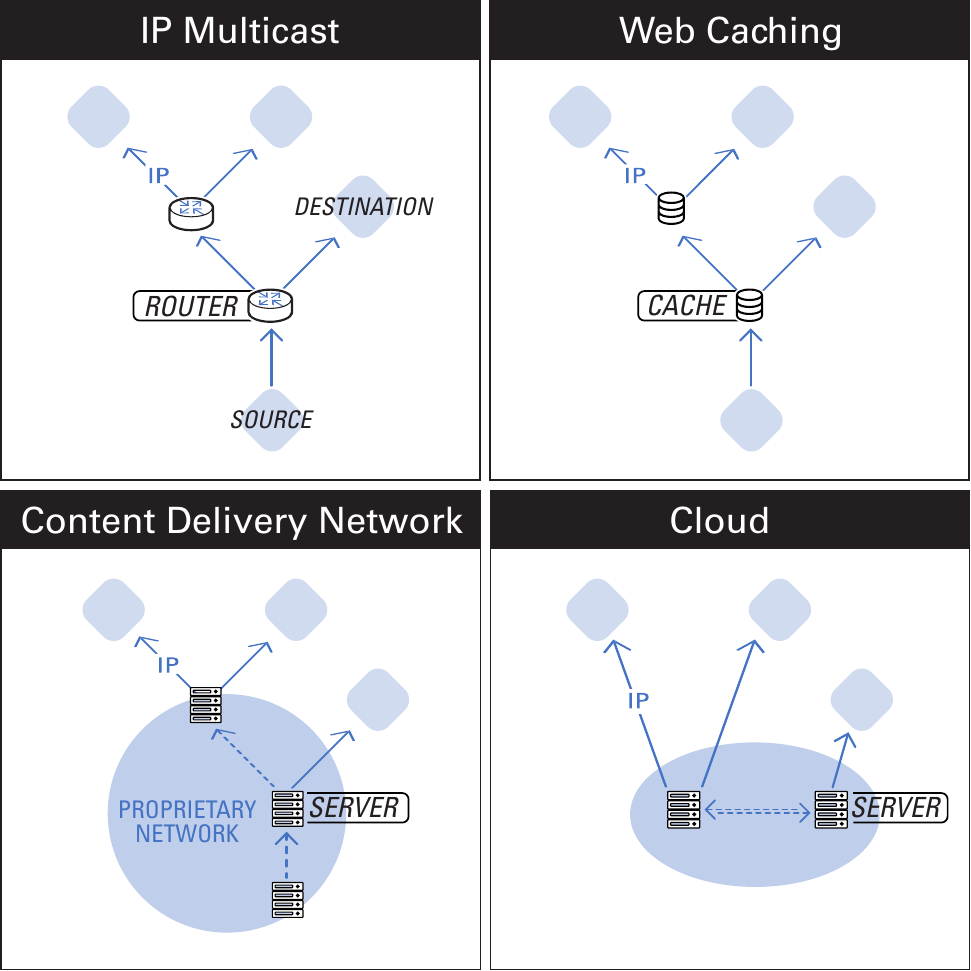}
\vspace*{-3ex}
\caption{Overlay Workarounds Addressing Point-to-Multipoint Distribution}
\vspace*{-5ex}
\label{fig:trees}
\end{figure}

During what might be called the ``Internet Convergence'' in the 1990`s,
the generality and scalability of the Internet's datagram delivery model gave
rise to the idea of using it to implement the convergence of broadcast,
telephony and data services~\cite{messerschmitt1996convergence}.
The emergence of unicast datagram delivery as the
only universal Internet service (discussed in Section~\ref{sec:background}) has
meant that the underlying capabilities of lower layer mechanisms to
utilize electromagnetic broadcast and to guarantee quality of service through
resource reservation are not accessible using Voice over IP and Streaming Media
over IP protocols. In spite of such limitations, the convenience and cost
benefits of convergence workarounds continue to dominate the commercial
development of these services.

But the absence of a universal point-to-multipoint communication mechanism within
the common Network layer of the Internet
left a large class applications without native support, and
this, in turn, has generated a whole series of
overlay workarounds (see Figure~\ref{fig:trees}). For instance, the
distribution of static Web pages (those that require only minimal rewriting of
stored HTML pages) can be viewed as a form of point-to-multipoint communication.
A browser cache uses moderate storage resources in the network endpoint to
capture the delivered Web page and associated metadata and minimal processing
to implement the cache policy and mechanism. A proxy cache uses larger scale
storage and has a greater processing load, which is supplied by a substantially
provisioned network intermediate node. The convergence of resources in Web
caches led to an architectural development in which application-specific
proxies are uploaded to the implementation of ``middlebox'' platforms which implements both
caching and general processing.

Web caching played a pivotal role in the expansion of the Web as a global data
distribution service during the period when intercontinental data links were
too expensive to allow unfettered access by academics.
A hierarchical system of large scale caches was developed and deployed in US
Research and Education Networks
\cite{Chankhunthod95ahierarchical,Wessels98icpand} and use of national caches
to access Web data across intercontinental links was made mandatory in many
countries.

In spite of its effectiveness in reducing the traffic loads due to delivery of
static Web pages, the popularity of intermediate caches has waned dramatically
in the past decade.
There are several reasons for this trend including:
\begin{itemize}
    \item The correctness of Web caching relies on  lifetime metadata being
      provided by origin servers which is often missing or inaccurate.
    \item The growth of dynamic Web applications means that many Web objects
      are not cachable.
    \item The lack of an accurate and universal mechanisms for reporting views
      interferes with the dominant business model of Web advertisers.
    \item Reliance on a complex cache infrastructure decreases the control by
      the implementer of a Web service over the Quality of Service experienced
      by customers.
\end{itemize}
Many of these factors stem from the implementation of Web caching on top of the
HTTP application protocol, albeit with some modifications to
increase control over intermediate and browser caches by origin Web servers.
Cache networks are an overlay which accesses Web services from the top of the
protocol stack and thus does not allow the degree of fine-grained control that
is required for seamless convergence.

An alternative approach is to start from the source, and to replicate the
functionality of the Web server on multiple network nodes.
Manual procedures for FTP mirroring led to automated mechanisms like
Netlib~\cite{DongarraNetlib08}, and high traffic Web sites gave rise to
sophisticated cluster and geographically distributed server replication
schemes~\cite{KirkpatrickIBMOlympics96,Beck98theinternet2}.

Commercial content Delivery Networks have approached the problem in a somewhat
different way, using
HTTP and streaming protocols for client access almost unchanged.
This is analogous to the way that online services (e.g. Compuserve and AOL) and
ISPs used telephone services.
CDNs have instead focused their innovation on the underpinnings of the Internet
in order to improve the effectiveness.
They use a combination of server side caching,
distributed file and database systems and complex streaming and synchronization
protocols implemented on private,  proprietary international networks of
application-specific servers.

CDN Web and DNS servers may be implemented as
applications processes, but by using lower layer Internet
mechanisms through now-commonplace layering violations (such as
topology-sensitive DNS resolution), they use knowledge of network topology and other low
level information that is intended to be encapsulated within the Network Layer
of the Internet architecture. Modern extensions to the Network
layer may allow CDN`s to be implemented without such
violation of layering, but at the expense of creating a ``fatter'' and less generic
Network layer (see Section~\ref{sec:scalability}).

Commercial  CDNs are thus a kind of Chimera, patched
together from proprietary components and standard, low level components of the
Internet. They create a
proprietary, specialized network with their own services as the spanning layer,
using the Internet as tools in their implementation and as a means of reaching
end users. This view is supported by the trend toward using private or non-scalable
mechanisms to implement internal communication among  centralized and
distributed CDN nodes. Since the currently emerging paradigm for
edge computing ``\ldots
extends the CDN concept by leveraging cloud computing
infrastructure\ldots...~\cite{satyanarayanan2017emergence},'' the
non-interoperable
nature of these overlay approaches undermines the coherence of
Internet-based information service ecosystems. As we argue below,
\emph{Exposed Buffer Processing} offers a more interoperable and unified
foundation for such next-generation ecosystems.




\section{Exposed Buffer Processing: An Approach to Interoperable Convergence}
\label{sec:ebp}

While operating system interfaces such as POSIX provide access to storage,
networking and computing services, they do so in ways that conform to the
traditional silos.
\begin{itemize}
\item File system calls do not have explicit access to general networking or
  computation resources.

\item The socket interface does not provide access to general storage and or
  computation resources.

\item The POSIX process management functions have only the minimal necessary
  overlap with storage and network functions (notably specifying an executable
  file image in the \texttt{exec()} system call).
\end{itemize}

However, the core resource that is used to implement all these silos is the
persistent memory or storage buffer.
\begin{itemize}
\item In {\it storage}, disk/SSD blocks or objects are used in the implementation of
  higher level file and database systems, along with RAM memory buffers that
  are used to improve performance, enable application/OS parallelism and allow
  for flexible exchange of data with other operating system data structures.

\item In {\it networking}, buffers are used at the endpoints for much the same
  reasons as storage, and are used at intermediate nodes to allow for
  asynchrony in the operation of store-and-forward packet networking.

\item In {\it computing}, memory pages make up process address spaces,  are also used
  to enable asynchrony in interprocess communication, and hold all other
  operating system data structures used in the implementation of functions on
  behalf of processes.
\end{itemize}

So although convergence of storage, networking and computation is possible through
conventional operating system interfaces using the generality of the user
process as a gateway between silos, a more interoperable approach is to expose a common abstraction of the
underlying resource that all of these high level silos operate on, namely
persistent storage blocks or memory buffers. We call this approach to
convergence {\em Exposed Buffer Processing}.

\subsection{Core EBP functionality}
Exposed Buffer Processing is a general architectural idea that can have different implementations.
The original implementation, which takes the form of the Logistical Networking
stack described in Section~\ref{subsec:logistical}, encapsulates the service as
remote procedure call over TCP.

\begin{itemize}

\item {\tt allocate} --
This call allocates storage capacity into which data can be stored.
It is possible for this allocation to be performed implicitly
(as part of the {\tt store} operation described below) or explicitly (as a
reservation of resources which subsequent {\tt store} operations can use.)
An allocation call specifies several parameters that limit resource utilization, such as duration.
One important attribute  of an allocation is a local name by which stored data is referenced in subsequent calls.

\item {{\tt store} and {\tt load}} --
These calls allow EBP clients to store data in an allocated buffer or to load data that has previously been stored.

\item {\tt transfer} --
This call transfers data between buffers specified by name on the same  or on different depots.

\item {\tt transform} --
This call applies a named operation to a set of buffers on a single depot, potentially transforming the data of one or more of them.
The operation name is local to the depot and must have been defined previous to
the transform call by a code upload operation which is specific to the
implementation of the depot.
Operations will be assigned names through a client-community process that ensures that names are used in a sufficiently consistent manner.
\end{itemize}

\subsection{Logistical Networking as EBP in Overlay}
\label{subsec:logistical}

Over the past 15 years the Logistical Networking
project~\cite{Plank:2001:MDS:613350.613637,Beck02anend-to-end,Beck03anend-to-end}
has worked to define an approach to Exposed Buffer Processing that is
implemented as an overlay on the Internet. An examination of the key components
of that implementation provides an EBP proof of concept:

\begin{itemize}

    \item \textbf{Internet Backplane Protocol (IBP)}:
    IBP is a generalized Internet-based storage service that is encapsulated
    as remote procedure call over TCP.
    IBP was designed to be simple, generic and limited following the example of the
    Internet Protocol (IP)~\cite{Saltzer:1984:EAS:357401.357402}.
    It is a best effort service, its byte array allocations are named only by long
    random server-chosen keys (capabilities) and represent leases whose duration and size are
    limited by the individual intermediate node (in analogy to the IP MTU).
    The intermediate node that implements IBP is called a \emph{depot}, and it is
    intended as a storage analog to IP routers.
    In many ways IBP is closer to a network implementation of \texttt{malloc()} than a
    conventional Internet storage service like FTP.
    In addition every IBP
    allocation is a lease of storage resources which can be limited in duration.
    The IBP depot has been implemented in both C and Java.

    \item \textbf{exNode}: Because IBP is such a limited service,
    the abstraction of an allocation that it
    supports does not have the expressiveness of the file abstraction that users
    typically expect of a high level data management system.
    The exNode is a data structure that holds the structural metadata
    required to compose IBP allocations into a file of very large extent, with
    replication across IBP depots, identified by their DNS name or IP
    address~\cite{Beck01logisticalcomputing}.
    The exNode can be thought of as an analog to the inode used in early Unix file
    system implementations.
    The exNode has both standard XML and JSON sequentializations.

    \item \textbf{Logistical Runtime System (LoRS)}: The exNode
    can be used as a file descriptor to implement standard file
    operations such as {\bf read} and {\bf write}.
    The Logisticsal Runtime System (LoRS) uses the exNode to implement efficient,
    robust and high performing data transfer operations.
    Some of the techniques used in the implementation of LoRS are comparable to
    those used in parallel and peer-to-peer file transfer protocols~\cite{Plank02algorithmsfor}.

    \item \textbf{Logistical Distribution Network (LoDN)}: While the
    exNode implements topological composition of IBP allocations to
    implement large distributed and replicated files, it does not deal with the
    temporal restrictions introduced by IBP's use of storage leases.
    LoDN is an active service which holds exNodes and applies storage allocation,
    lease renewal and data movement operations as required to maintain policy
    objectives set by end users through a declarative language and manageable by an
    intuitive human interface.

    \item \textbf{Network Functional Unit (NFU)}:The NFU was introduced as a
    means to allow simple, generic and limited in-situ
    operations by a depot to data stored in its IBP allocations.
    The NFU has been used in numerous experimental deployments, and has been shown
    to enable robust fault tolerance and high performance in a wide variety of
    applications~\cite{Liu06DynamicCosched,Beck_scalabledistributed:2007,Liu:Dissertation:2008}.
    However, the middleware stack that supports such experimentation has never
    been fully integrated with the packaged versions of LoRS and LoDN or the Data
    Logistics Toolkit (discussed below), and so the
    NFU has never been used in a persistent large scale deployment.
\end{itemize}

\subsection{``Packetization'' of Storage and Processing}

One way to characterize EBP`s simple, generic, and limited design philosophy for
the abstractions of common spanning layer services is to say that
it extends the idea of ``packetization'' from the domain of networking,
where it has proved so remarkably successful, to the domains storage/memory and
processing as well. Unfortunately, this
contradicts the impulses many service architects who have
historically relied on the more complex,
specialized and virtually unbounded services.
The relevant contrasts between packet-based and
circuit-based approaches are familiar and clear in realm of \textbf{Networking}:

\begin{itemize}
    \item \textbf{Size}: Circuit-based networks allow an unbounded amount of data to pass
    over a persistent circuit, in analogy to an electrical connection,
    masking the underlying digital implementation in terms of MTU-limited
    packets.
    The Internet exposed the MTU and required endpoints to concatenate
    packets into streams.

    \item \textbf{Failure}: Circuit-based networks provide Quality-of-Service (QoS)
    guarantees sufficient to enable application developers to be protected from
    occasional communication faults but which fail catastrophically when protection is impossible.
    The Internet exposed the possibility of failure
    by exporting a best effort service, requiring endpoints to
    detect and respond to failures.

    \item \textbf{Locality Independence}: Circuit-based networks can allocate resources and maintain state along a
    specific path from sender to receiver, helping to ensure fast forwarding
    and providing a stable platform for implementation of auxilliary services.
    The Internet allows every packet in a connected flow to be forwarded along a
    different path, putting the burden for maintaining stability on the packet
    routing scheme and ruling out connected services that require the
    maintenance of state, but enabling great resilience in the face of failures
    and changes in topology.
\end{itemize}

The similarities between between networking and storage make the
the analogous set of contrasts relatively easy to work out
for the realm of \textbf{Storage}:

\begin{itemize}

    \item \textbf{Size}: File-based models of storage allow a very large amount of data
    (assumed by many applications to be virtually unbounded) to be stored as a single linear
    data extent. Logistical Networking (i.e., EBP in overlay) exposes a maximum
    storage allocation size imposed by the
    storage resource (analogous to the Internet Protocol's MTU) requiring
    endpoints to explicitly concatenate allocations into files.

    \item \textbf{Failure}: File and database systems provide reliability guarantees sufficient to enable
    application developers be protected from occassional storage faults but which fail catastrophically when protection is impossible.
    Logistical Networking exposes a simple failure model (faulty write  operations
    terminate with unknown output state) and by exporting
    a best effort service, requiring endpoints to explicitly detect and respond
    to failures.

    \item \textbf{Locality Independence}: File-based models of storage can allocate
    resources and maintain state on a well-connected ``site'' to manage fault
    tolerance and replication in terms of where ``copies'' reside.
    Logistical Networking allows every allocation comprising a file to be managed
    independently, potentially spreading them across topologically seperated
    nodes, moving and storing data  on a fine-grained basis as called for by
    applications (e.g., data streaming).

\end{itemize}

Finally
on which it operates,
an analogous set of contrasts
applies to the realm of \textbf{Computation}:

\begin{itemize}
    \item \textbf{Size}: Process-based computation allows an unbounded amount of processing to be
    performed by a set of one or more closely-coupled threads.
    The Network Functional Unit (i.e., EBP in overlay) exposes a unit of
    processing that can be limited in
    many resource dimensions, including CPU cycles
    consumed, RAM allocated during execution and I/O activity performed,
    requiring a runtime system to concatenate limited resources to create an
    unbounded virtual execution model.

    \item \textbf{Failure}: Process-based computation provides QoS guarantees sufficient to enable
    application developers to either overcome occasional processing faults or to
    fail catastrophically when they are detected.
    The NFU exposes a simple failure model (faulty operations
    terminate with unknown state for write-accessible storage) and exports a
    best effort service, requiring endpoints to explicitly detect and respond
    to failures.

    \item \textbf{Locality Independence}: Process-based computation can allocate resources and maintain state on a set of
    well-connected processors, enabling successive time slices to execute
    sequentially in a manner that leverages continuity of operating system  and
    application data state.
    The NFU allows every allocation comprising a process to be
    managed independently, potentially moving them and the memory/storage
    allocations that comprise the state of supervisory and application data
    state as required (eg fault tolerance and load balancing).
\end{itemize}

\subsection{EBP over Packet}
Currently, the IBP protocol is encapsulated over TCP. Using TCP as a substrate
offers several benefits and simplifies the engineering effort for the EBP
developer, but this convenience comes at a cost. In particular, many design
decisions baked into TCP were made to serve the needs of public wide-area
networks that must support competing large data transfers that implement fair
contention and flow control to facilitate resource sharing.

Not all of these features are necessary for EBP to work efficiently, and in
fact, some are detrimental in certain contexts. For example, strict in-order
delivery of network packets can inhibit parallelism, and slow start imposes
unnecessary latency when an EBP operation involves delivery of a number of
packets that is too small to cause congestion.
For this reason it is useful to consider a possible implementation of EBP as it
would be encapsulated more directly over a packet transport, be that IP v4 or
v6, Ethernet or another Link Layer service.  Considering such a primitive
encapsulation lays bare the relationship between the scale of communication,
storage and processing, and how these can be reconstructed in a converged way
if they are expressed as the aggregation of limited operations.

\begin{itemize}
\item \textbf{Operation sequencing}: TCP delivers packets in order, which means
that if a packet arrives earlier than it is expected, it will not be
delivered to the upper layer until the ``missing'' packets arrive. When
these packets contain parts of a data buffer, then it is essential that
they are assembled in the correct order. However, requests for multiple EBP
operations--- which in a packet encapsulation would be transmitted as
multiple packets--- might depend on one another or they might be completely
independent. When two independent operations are sequenced in a TCP stream,
the second may arrive at its destination before the first but not be
delivered for processing, causing delays and limiting parallelism. To
satisfy the correctness requirements of operations that {\em do} depend on
one another specific operation packets can be tagged and {\em necessary}
order imposed. There are trade-offs between epressiveness and cost (in
terms of bits used by different tagging mechanisms for storing the tag)
which can be addressed using techniques developed for dynamically scheduled
systems for expressing dependencies between tasks as DAGs.

\item \textbf{Retransmission}: TCP provides mechanisms for retransmitting
packets that are not acknowledged within a timeout using exponential backoff
to avoid congestion. The approach taken by TCP is geared to maintaining
in-order delivery and to keeping the avoidance of congestion paramount in
the use of shared wide area networks. In some EBP scenarios, time-critical
operations could retransmit packets  much more aggressively to minimize the
possibility of all packets being lost. However, in order to maintain
correctness in the presence of non-idempotent EBP operations (such as those
with side-effects) it will be necessary to implement bookkeeping at the
destination to avoid {\it stuttering}.

\item \textbf{Flow control}: TCP provides flow control that is tied to
detection of lost packets at the expense of throttling high bandwidth
transfers. In cases where flow control is needed, it will have to be
implemented at a higher layer.
\end{itemize}

Since EBP offers several services simultaneously---storage, networking, and
computation---there are several aspects of QoS that can be explored.

  \begin{itemize}
  \item \textbf{Allocation of Bandwidth/Storage/Computing.}
  The most fundamental functionality that EBP provides is allocation of
  resources. Especially in the scenario where EBP is implemented over IP,
  and thus the fair contention safeguards of TCP are bypassed, a client
  application could request to allocate the whole bandwidth capacity of a
  network route. In this scenario competing allocations would have to be
  declined. Similarly, one can envision multiple allocations that only
  require part of the network's capacity to be satisfied simultaneously. In
  the same spirit, and using similar bookkeeping mechanisms, we could
  implement allocation of storage and computing resources on the nodes.

  \item \textbf{\textit{Hard} versus \textit{soft} allocations.}
  As an additional QoS feature we can provide multiple levels of allocation
  ``hardness'' which would trade the level of guarantee for the amount of
  resources available. At least three levels of hardness can easily be
  envisioned: A) best-effort clients, which make no allocation (and thus
  receive no QoS guarantee) can make unlimited attempts to allocate
  resources; B) clients which make \textit{soft} allocations can allocate a
  resource before it is used are first to be preempted when that resource
  is exhausted due to competing system activities or overbooking of hard
  allocations (see below). Soft allocations can result in the denial of
  best-effort clients even when resources are not exhausted; and C) clients
  which make \textit{hard} allocations can preempt all other types of
  clients and are last to be preempted is case of total resource
  exhaustion.

  \item \textbf{Statistical overbooking,} can be used to allow resource
  reservation to exceed available resources to take advantage of
  underutilization of reserved resources. Overbooking can be managed
  through a multi-tiered schema of allocation ``hardness'' as described
  above.

\end{itemize}

\subsection{EBP Below the Network Layer}


The argument for creating a  converged layer to support global distributed services is compelling.
The need for distributed systems to have access to and control over low layer
network characteristics including topology and performance is clear in the
steps that have been taken to work around the stricture that forbids such
direct access in the Internet architecture.

We propose the creation of a platform based on a common service similar to IBP
but which models the networking capabilities of the Link Layer.
We use the term Exposed Buffer Processing for this as-yet-unrealized service.
The central idea of this paper is that the appropriate platform for the
creation of distributed systems is some form of EBP.
We emphasize that EBP need not follow the design of IBP, as long as it takes
appropriate account of the Deployment Scalability Tradeoff.
We offer experience with IBP as an overlay form of EBP for the
consideration of the community.

\section{Applications of EBP}
\label{sec:applications}

%

\subsection{Scientific Content Delivery}

Dissemination of data is one of the fundamental challenges of modern
experimental and observational science.
There is a general move toward the open sharing of raw data sets, enabling
replication of analyses, cross-cutting studies, innovative reexamination of
previously collected data and historical examination of collection and analysis
techniques~\cite{reichman2011challenges,kitchin2014data}.
In many case the data collected is large and observation is continuous, as in
remote  data from satellites and other sensors~\cite{board2014landsat},
experiments such as the Large Hadron Collider~\cite{bird2011computing}, or
broad harvesting of multimedia content~\cite{breeding2003building}.
The resources required to make such data streams instantaneously and
persistently available can exceed the centralized capabilities of  institutions
or government agencies.

Commercial CDN or Cloud solutions may be too expensive, and may not adequately
serve the entire global user community (see discussion of the Digital Divide
below) and may not adequately support the publication by users of secondary
data products resulting from their processing of raw data.
However, the ICT resources required to address such problems may be affordable,
and the community of user institutions may be capable of hosting them in a
distributed manner.
Using shared EBP infrastructure, we can build a distributed, federated content
management system using the resources of the content provider and user
communities

\subsection{Digital Divide and Disasters}

Modern network services take full advantage of the strong assumptions that can
be made about the implementation of the Internet in the industrial world.
It is common for services to rely on continual low-latency datagram delivery,
always-connected servers, stable and uninterrupted datagram routing paths and
high bandwidth connectivity to take just a few examples.
Services implemented at Cloud Computing centers are among those that place
great demands on the Internet backbone and ``last mile'' connectivity to edge
networks.

Many services can be decomposed into synchronous and asynchronous components,
and different ``Data Logistics'' strategies applied to each part~\cite{datalogistics}.
Techniques used in Content Delivery Networks, including caching and prestaging,
can be applied on a fine-grained and even per-client basis.
It is sometimes the case that the entire service can be implemented using edge
resources. In other cases there is a component that can only be implemented
using synchronous end-to-end datagram delivery across the backbone, but which
requires only low bandwidth.
In some cases analysis of the application combined with
reconsideration of the truely necessary characteristics of the  service
delivered to the end-user can reduce the need for high quality synchronous
connectivity to the vanishing point.
In a sense, reliance on strong network assumptions is often used to trade off
unnecessary reliace on excellent network infrastructure for ease of
development. This is a useful strategy for those who can afford and support the
necessary infrastructure.

Today, some environments cannot support strong network assumptions, even when
local IT resources are available.
Examples are communities isolated through geography, economic (poverty,
discrimination), political circumstances (famine, war), or social factors.
Disasters create environments where infrastructure is disrupted even in the
most advanced societies.
The recent response of modern network technologists has been to bring fixed or
mobile wireless technology (satellite, 4G) into remote locations and to the
scene of disasters or to create complex wireless infrastructures based on
continuous aviation drones such as Google's baloon-based project
Loon
and Facebook's drone-based project
Aquila.
In contrast, using a mix of interoperable heterogeneous synchronous and
asynchronous data transport integrated into a flexible platform to support a
variety of distributed applications can be cheap, robust and
easily deployed.

\subsection{Big Data and Edge Processing}

One of the inexorable trends in the collection of data is the emergence of
large scale online sensors and instrument that produce data that must be
subjected to volume-reducing processing before it can be passed over the
network. Growing trends in sensor networks, the Internet of Things, and Smart
Cities will severely exacerbate this problem~\cite{banerjee2013final}.
The historical approach has been to send all such data to computation centers that
are either self-contained or connected to their peers through heroic networking
that may be private or even proprietary. This is no longer sufficient to
address the total size of data, its globally distributed generation and
consumption of data that we see today~\cite{chen2014big}.  An alternative
possible using EBP is to apply limited edge processing on the in the edge
network using a converged infrastructure that can also store and transport
data.

\subsection{In-locus Data Analysis}

Data Analytics (DA) has emerged as a new paradigm for understanding unreliable and
varying environments.  Going beyond logging, reporting, and
thresholding, DA can perform meaningful analyses of large scale data sources that are networked
through dynamic and distributed infrastructure. (The stage before batch or
streaming analytics take place is often called ``data assimilation''.)
DA is capable of extracting
latent knowledge and providing insight from field sensors, computational units,
and large mobile networks.  Of course the number of these data sources
and the corresponding ingest rates are growing dramatically because of increased edge
hardware capability (resolution and sampling rate) and hybridization
(multi-messenger and multi-sensor data acquisition). These factors require new algorithmic
approaches that closely integrate the network, I/O, and computational software
stacks to lower the overheads and provide non-trivial data metrics at the edge.
Fortunately, the Applied Mathematics and
Machine Learning communities have recently produced innovations in the field of
approximate and/or randomized algorithms, which combines new methods for matrix
approximation via random sampling, that are perfectly positioned to fill this
role.

Recent work in
randomized and approximate algorithms~\cite{avron2010blendenpik,Drineas:2016:randnla},
which attach a probabilistic measure to their results, improves the fit of such methods
for inherently unstable and constantly changing distributed environments.
In fact, there are many
statistical techniques in the Randomized Linear Algebra class of algorithms
that lend themselves perfectly to utilization in the converged
approach of in-locus computing, e.g., by using IBP's best effort Network
Functional Unit operations as discussed in Section~\ref{subsec:logistical}. Such NFU operations can respond
algorithmically to assimilate the inherent failures that
naturally occur in a widely distributed system at the scale that we target. The
iterative nature of most approximate methods allows us to incorporate erroneous
responses from a sensor or a network transmission and gradually remove the
malformed data from the multidimensional subspace that is being worked on.
Similarly, an intermittent lack of response from a sensor or a network element
may naturally be incorporated as a sampling and selection operator that is
triggered by a system-reported event as opposed to the classical method that
uses a pseudo random number generator (PRNG) as an unbiased projector or selector.
Also, the probabilistic nature of
the approximate algorithms allows us to weigh the data sources based on their
history of reliable responses and the quality of the data they delivered (if a
measure of quality can be obtained, from, for example, a duplicate sensor).
High quality sensors and network connections will, over time, gain larger
weights, in turn rendering them highly probable to be approximately correct as
envisioned by the Probably Approximately Correct (PAC) learning
framework~\cite{Valiant:1984:prob_approx_correct}.

The fundamental operators of randomized methods are \emph{selection} and
\emph{projection}. They may be used in combination or individually, depending on
the need. In our approach, we use the transmission errors as a form of
selection, while projections would mostly be constructed to incorporate
knowledge about the state of the system. In statistical parlance, we strive to
obtain unbiased sampling of the incoming data. As long as the sample is
representative and limited in size, we are able to process the data at the
edge or afford the bandwidth to send closer to the core where more computing
power resides. The prior distribution is constructed from known pieces of
information, such as hardware specification of the sensors and their reporting
frequency. Over time, we may be able to form a more useful posterior
distribution that is informed by changes in the infrastructure deployed in the
field and in the surroundings that are being monitored by the sensors.  In
order to achieve this goal, the integration of the algorithmic methods and the
system stack needs to occur in novel ways. Contrast this with a rather
idealistic notions of the prevailing fault tolerant paradigms that tend to
assume that errors occur discretely in the midst of reliable computing
periods. It is assumed that error correction can restore the corrupted
data through some form of redundant state management (e.g. checkpoints
or error correcting codes) to maintain the logical invariant of the
data being unaltered. We move away from these idealizations and incorporate
the probability of errors directly into the probability of the answer being
correct, using biased sampling to isolate valuable data that has high
probability of being representative of the state of the system. The bias in our
approach is guided by the information obtained either statically, before
execution, and dynamically as the computation progresses.

\section{Conclusions}
\label{conclusions}


In this paper, we have argued that interoperable convergence of storage,
networking and processing is necessary in building a platform to support
distributed systems which exhibits deployment scalability, and that the most
effective implementation is a form of Exposed Buffer Processing at a layer
below that which implements the Internet.
Our argument rests on practical historical examples of the problems caused by
the Internet's lack of generalized state management and an argument based on a partially
formalized design methodology that the spanning layer of any converged
infrastructure must be simple, generic and limited.


\section*{Acknowledgements}

The ideas in this paper were influenced by many spirited discussions with
Martin Swany on the integration of storage and processing with scalable
networking, and by recent conversations with Glenn Ricart on the definition and
justification of interoperable convergence.  The concept of ``exposed buffer
protocol/processing'' was coined during discussions between Swany and  Beck,
although its best definition and implementation are still subject to debate.
The authors are also indebted to David Rogers for his professional rendering of
the artwork in this any many other papers and presentations, and to Chris
Brumgard for his helpful comments.





\bibliographystyle{abbrv}
\bibliography{hglass,ebpbib,randla}

\begin{thebibliography}{10}

\bibitem{RealE2e}
Will the real end-to-end argument please stand up?
\newblock http://mercury.lcs.mit.edu/~jnc/tech/end{\textunderscore}end.html.

\bibitem{uuxPOSIX}
Uux(1p) posix programmer's manual, 2013.

\bibitem{Anderson05overcomingbarriers}
T.~Anderson, L.~Peterson, S.~Shenker, and J.~Turner.
\newblock Overcoming the internet impasse through virtualization.
\newblock {\em Computer}, 38(4):34--41, April 2005.

\bibitem{avron2010blendenpik}
H.~Avron, P.~Maymounkov, and S.~Toledo.
\newblock Blendenpik: Supercharging {LAPACK}'s least-squares solver.
\newblock {\em SIAM Journal on Scientific Computing}, 32(3):1217--1236, 2010.

\bibitem{banerjee2013final}
S.~Banerjee and D.~O. Wu.
\newblock {\em Final report from the NSF Workshop on Future Directions in
  Wireless Networking}.
\newblock National Science Foundation, 2013.

\bibitem{beck_hourglass_2018}
M.~Beck.
\newblock On the {Hourglass} {Model}, {End}-to-{End} {Arguments}, and
  {Deployment} {Scalability}.
\newblock {\em Communications of the ACM}, to appear, 2018.

\bibitem{Beck01logisticalcomputing}
M.~Beck, D.~Arnold, R.~Bassi, F.~Berman, H.~Casanova, T.~Moore, G.~Obertelli,
  J.~Plank, M.~Swany, S.~Vadhiyar, and R.~Wolski.
\newblock Logistical computing and internetworking: Middleware for the use of
  storage in communication.
\newblock In {\em In 3rd Annual International Workshop on Active Middleware
  Services (AMS)}, 2001.

\bibitem{Beck_scalabledistributed:2007}
M.~Beck, H.~Liu, J.~Huang, and T.~Moore.
\newblock Scalable distributed execution environment for large data
  visualization.
\newblock {\em IEEE Explorer}, Nov. 2007.

\bibitem{Beck98theinternet2}
M.~Beck and T.~Moore.
\newblock The {Internet2} distributed storage infrastructure project: An
  architecture for internet content channels.
\newblock In {\em Computer Networking and ISDN Systems}, pages 2141--2148,
  1998.

\bibitem{Beck02anend-to-end}
M.~Beck, T.~Moore, and J.~S. Plank.
\newblock An end-to-end approach to globally scalable network storage.
\newblock In {\em In ACM SIGCOMM 2002}, 2002.

\bibitem{Beck03anend-to-end}
M.~Beck, T.~Moore, and J.~S. Plank.
\newblock An end-to-end approach to globally scalable programmable networking.
\newblock In {\em Future Directions in Network Architecture}, pages 328--339.
  ACM Press, 2003.

\bibitem{bird2011computing}
I.~Bird.
\newblock Computing for the {Large} {Hadron} {Collider}.
\newblock {\em Annual Review of Nuclear and Particle Science}, 61:99--118,
  2011.

\bibitem{board2014landsat}
S.~S. Board, N.~R. Council, et~al.
\newblock {\em Landsat and Beyond: Sustaining and Enhancing the Nation's Land
  Imaging Program}.
\newblock National Academies Press, 2014.

\bibitem{breeding2003building}
M.~Breeding.
\newblock Building a digital library of television news.
\newblock {\em Computers in libraries}, 23(6):47--49, 2003.

\bibitem{BuyyaCDN08}
R.~Buyya, M.~Pathan, and A.~Vakali, editors.
\newblock {\em Content Delivery Networks}.
\newblock Springer, 2008.

\bibitem{rfc3234}
B.~Carpenter and S.~Brim.
\newblock Middleboxes: Taxonomy and issues.
\newblock RFC 3234, Feb. 2002.
\newblock Network Working Group.

\bibitem{Chankhunthod95ahierarchical}
A.~Chankhunthod, P.~B. Danzig, C.~Neerdaels, M.~F. Schwartz, and K.~J. Worrell.
\newblock A hierarchical {Internet} object cache.
\newblock In {\em IN PROCEEDINGS OF THE 1996 USENIX TECHNICAL CONFERENCE},
  pages 153--163, 1995.

\bibitem{chen2014big}
M.~Chen, S.~Mao, and Y.~Liu.
\newblock Big data: A survey.
\newblock {\em Mobile Networks and Applications}, 19(2):171--209, 2014.

\bibitem{Chun:2003:POT:956993.956995}
B.~Chun, D.~Culler, T.~Roscoe, A.~Bavier, L.~Peterson, M.~Wawrzoniak, and
  M.~Bowman.
\newblock Planetlab: An overlay testbed for broad-coverage services.
\newblock {\em SIGCOMM Comput. Commun. Rev.}, 33(3):3--12, July 2003.

\bibitem{clark1995interoperation}
D.~D. Clark.
\newblock Interoperation, open interfaces, and protocol architecture.
\newblock {\em The Unpredictable Certainty: White Papers}, (2):133--144, 1995.

\bibitem{david2007path}
P.~A. David.
\newblock Path dependence: a foundational concept for historical social
  science.
\newblock {\em Cliometrica}, 1(2):91--114, 2007.

\bibitem{Dean:2008:MSD:1327452.1327492}
J.~Dean and S.~Ghemawat.
\newblock Mapreduce: Simplified data processing on large clusters.
\newblock {\em Commun. ACM}, 51(1):107--113, Jan. 2008.

\bibitem{DongarraNetlib08}
J.~Dongarra, G.~H. Golub, E.~Grosse, C.~Moler, and K.~Moore.
\newblock Netlib and {NA-Net}: Building a scientific computing community.
\newblock {\em IEEE Annals of the History of Computing}, 30:30--41, Apr. 2008.

\bibitem{Drineas:2016:randnla}
P.~Drineas and M.~W. Mahoney.
\newblock {RandNLA}: Randomized numerical linear algebra.
\newblock {\em Communications of the ACM}, 59(6):80--90, 2016.

\bibitem{foster1999grid}
I.~Foster and C.~Kesselman.
\newblock {\em The Grid: Blueprint for a New Computing Infrastructure}, pages
  47--48.
\newblock Advanced computing. Computer systems design. Morgan Kaufmann
  Publishers, 1999.

\bibitem{KirkpatrickIBMOlympics96}
D.~Kirkpatrick.
\newblock {IBM's} olympic fiasco department of groundless optimism.
\newblock {\em Fortune Magazine}, September 9 1996.

\bibitem{kitchin2014data}
R.~Kitchin.
\newblock {\em The data revolution: Big data, open data, data infrastructures
  and their consequences}.
\newblock Sage, 2014.

\bibitem{Liu:Dissertation:2008}
H.~Liu.
\newblock {\em Scalable, Data-Intensive Network Computation}.
\newblock PhD thesis, University of Tennessee, Knoxville, 2008.

\bibitem{Liu06DynamicCosched}
H.~Liu, M.~Beck, and J.~Huang.
\newblock Dynamic co-scheduling of distributed computation and replication.
\newblock In {\em IEEE International Symposium on Cluster Computing and the
  Grid}, May 2006.

\bibitem{GENIMcGeer16}
R.~McGeer, M.~Berman, C.~Elliott, and R.~Ricci, editors.
\newblock {\em The GENI Book}.
\newblock Springer, 2016.

\bibitem{messerschmitt1996convergence}
D.~G. Messerschmitt.
\newblock The convergence of telecommunications and computing: What are the
  implications today?
\newblock {\em Proceedings of the IEEE}, 84(8):1167--1186, 1996.

\bibitem{mynatt2017national}
E.~Mynatt, J.~Clark, G.~Hager, D.~Lopresti, G.~Morrisett, K.~Nahrstedt,
  G.~Pappas, S.~Patel, J.~Rexford, H.~Wright, et~al.
\newblock A national research agenda for intelligent infrastructure.
\newblock {\em arXiv preprint arXiv:1705.01920}, 2017.

\bibitem{nahrstedt2017city}
K.~Nahrstedt, C.~G. Cassandras, and C.~Catlett.
\newblock City-scale intelligent systems and platforms.
\newblock {\em arXiv preprint arXiv:1705.01990}, 2017.

\bibitem{nitrd_scc}
Networking, I.~T. Research, and D.~N. Program.
\newblock Smart and {C}onnected {C}ities {F}ramework, 2015.
\newblock \url{https://www.nitrd.gov/sccc/materials/scccframework.pdf}.

\bibitem{Nygren10theakamai}
E.~Nygren, R.~K. Sitaraman, and J.~Sun.
\newblock The {Akamai} network: A platform for high-performance {Internet}
  applications.
\newblock {\em SIGOPS Oper. Syst. Rev}, 2010.

\bibitem{Peterson:2011:CNF:2012058}
L.~L. Peterson and B.~S. Davie.
\newblock {\em Computer Networks, Fifth Edition: A Systems Approach}.
\newblock Morgan Kaufmann Publishers Inc., San Francisco, CA, USA, 5th edition,
  2011.

\bibitem{Plank02algorithmsfor}
J.~S. Plank, S.~Atchley, Y.~Ding, and M.~Beck.
\newblock Algorithms for high performance, wide-area, distributed file
  downloads.
\newblock Technical report, LETTERS, 2002.

\bibitem{Plank:2001:MDS:613350.613637}
J.~S. Plank, A.~Bassi, M.~Beck, T.~Moore, D.~M. Swany, and R.~Wolski.
\newblock Managing data storage in the network.
\newblock {\em IEEE Internet Computing}, 5(5):50--58, Sept. 2001.

\bibitem{reichman2011challenges}
O.~J. Reichman, M.~B. Jones, and M.~P. Schildhauer.
\newblock Challenges and opportunities of open data in ecology.
\newblock {\em Science}, 331(6018):703--705, 2011.

\bibitem{Ritchie74theunix}
D.~M. Ritchie and K.~Thompson.
\newblock The {U}nix time-sharing system.
\newblock {\em Communications of the ACM}, 17:365--375, 1974.

\bibitem{Saltzer:1984:EAS:357401.357402}
J.~H. Saltzer, D.~P. Reed, and D.~D. Clark.
\newblock End-to-end arguments in system design.
\newblock {\em ACM Trans. Comput. Syst.}, 2(4):277--288, Nov. 1984.

\bibitem{satyanarayanan2017emergence}
M.~Satyanarayanan.
\newblock The emergence of edge computing.
\newblock {\em Computer}, 50(1):30--39, 2017.

\bibitem{Tennenhouse96towardsan}
D.~L. Tennenhouse and D.~J. Wetherall.
\newblock Towards an active network architecture.
\newblock {\em Computer Communication Review}, 26:5--18, 1996.

\bibitem{Valiant:1984:prob_approx_correct}
L.~G. Valiant.
\newblock A theory of the learnable.
\newblock {\em Communications of the ACM}, 27:1134--1142, 1984.

\bibitem{Wessels98icpand}
D.~Wessels and k~claffy.
\newblock {ICP} and the {Squid} {Web} cache.
\newblock {\em IEEE JOURNAL ON SELECTED AREAS IN COMMUNICATION}, 16:345--357,
  1998.

\bibitem{datalogistics}
{Wikipedia contributors}.
\newblock Information logistics, 2017.
\newblock [Online; accessed 15-September-2018].

\end{thebibliography}

\end{document}